\PassOptionsToPackage{unicode}{hyperref}
\PassOptionsToPackage{hyphens}{url}
\PassOptionsToPackage{dvipsnames,svgnames,x11names}{xcolor}
\documentclass[
]{article}
\usepackage[a4paper, total={5.85in, 9in}]{geometry}
\usepackage{amsmath,amssymb}
\usepackage{amsbsy}
\usepackage{lmodern}
\usepackage{iftex}
\usepackage{float}
\ifPDFTeX
  \usepackage[T1]{fontenc}
  \usepackage[utf8]{inputenc}
  \usepackage{textcomp} % provide euro and other symbols
\else % if luatex or xetex
  \usepackage{unicode-math}
  \defaultfontfeatures{Scale=MatchLowercase}
  \defaultfontfeatures[\rmfamily]{Ligatures=TeX,Scale=1}
\fi
\IfFileExists{upquote.sty}{\usepackage{upquote}}{}
\IfFileExists{microtype.sty}{% use microtype if available
  \usepackage[]{microtype}
  \UseMicrotypeSet[protrusion]{basicmath} % disable protrusion for tt fonts
}{}
\makeatletter
\@ifundefined{KOMAClassName}{% if non-KOMA class
  \IfFileExists{parskip.sty}{%
    \usepackage{parskip}
  }{% else
    \setlength{\parindent}{0pt}
    \setlength{\parskip}{6pt plus 2pt minus 1pt}}
}{% if KOMA class
  \KOMAoptions{parskip=half}}
\makeatother
\usepackage{xcolor}
\usepackage{longtable,booktabs,array}
\usepackage{calc} % for calculating minipage widths
\usepackage{etoolbox}
\makeatletter
\patchcmd\longtable{\par}{\if@noskipsec\mbox{}\fi\par}{}{}
\makeatother
\IfFileExists{footnotehyper.sty}{\usepackage{footnotehyper}}{\usepackage{footnote}}
\makesavenoteenv{longtable}
\usepackage{graphicx}
\makeatletter
\def\maxwidth{\ifdim\Gin@nat@width>\linewidth\linewidth\else\Gin@nat@width\fi}
\def\maxheight{\ifdim\Gin@nat@height>\textheight\textheight\else\Gin@nat@height\fi}
\makeatother
\setkeys{Gin}{width=\maxwidth,height=\maxheight,keepaspectratio}
\makeatletter
\def\fps@figure{htbp}
\makeatother
\NewDocumentCommand\citeproctext{}{}
\NewDocumentCommand\citeproc{mm}{%
  \begingroup\def\citeproctext{#2}\cite{#1}\endgroup}
\makeatletter
 \let\@cite@ofmt\@firstofone
 \def\@biblabel#1{}
 \def\@cite#1#2{{#1\if@tempswa , #2\fi}}
\makeatother
\newlength{\cslhangindent}
\newlength{\csllabelwidth}
\newenvironment{CSLReferences}[2] % #1 hanging-indent, #2 entry-spacing
 {\begin{list}{}{%
  \setlength{\itemindent}{0pt}
  \setlength{\leftmargin}{0pt}
  \setlength{\parsep}{0pt}
  \ifodd #1
   \setlength{\leftmargin}{\cslhangindent}\renewcommand*{\arraystretch}{1.3}
   \setlength{\itemindent}{-1\cslhangindent}
  \fi
  \setlength{\itemsep}{#2\baselineskip}}}
 {\end{list}}
\usepackage{calc}

\ifLuaTeX
\usepackage[bidi=basic]{babel}
\else
\usepackage[bidi=default]{babel}
\fi
\babelprovide[main,import]{american}
\def\languageshorthands#1{}
\ifLuaTeX
  \usepackage{selnolig}  % disable illegal ligatures
\fi
\IfFileExists{bookmark.sty}{\usepackage{bookmark}}{\usepackage{hyperref}}
\IfFileExists{xurl.sty}{\usepackage{xurl}}{} % add URL line breaks if available
\hypersetup{
  pdftitle={PyBOP: A Python package for battery model optimisation and
parameterisation},
  pdfauthor={Brady Planden, Nicola E. Courtier, Martin Robinson, Agriya
Khetarpal, Ferran Brosa Planella, David A. Howey},
  pdflang={en-US},
  colorlinks=true,
  linkcolor={Maroon},
  filecolor={Maroon},
  citecolor={Blue},
  urlcolor={Blue},
  pdfcreator={LaTeX via pandoc}}

\title{PyBOP: A Python package for battery model optimisation and
parameterisation}

\definecolor{c53baa1}{RGB}{83,186,161}
\definecolor{c202826}{RGB}{32,40,38}

\usepackage[affil-it]{authblk}
\usepackage{orcidlink}
\author[1%
  ]{Brady Planden%
    \,\orcidlink{0000-0002-1082-9125}\,%
    }
\author[1,2%
  ]{Nicola E. Courtier%
    \,\orcidlink{0000-0002-5714-1096}\,%
    }
\author[3%
  ]{Martin Robinson%
    \,\orcidlink{0000-0002-1572-6782}\,%
    }
\author[4%
  ]{Agriya Khetarpal%
    \,\orcidlink{0000-0002-1112-1786}\,%
    }
\author[2,5%
  ]{Ferran Brosa Planella%
    \,\orcidlink{0000-0001-6363-2812}\,%
    }
\author[1,2%\ensuremath\mathparagraph
  ]{David A. Howey%
    \,\orcidlink{0000-0002-0620-3955}\,%
    }

\affil[1]{Department of Engineering Science, University of Oxford,
Oxford, UK%
  }
\affil[2]{The Faraday Institution, Harwell Campus, Didcot, UK%
  }
\affil[3]{Research Software Engineering Group, University of Oxford,
Oxford, UK%
  }
\affil[4]{Quansight Labs%
  }
\affil[5]{Mathematics Institute, University of Warwick, Coventry, UK%
  }
\date{20 December 2024}

\begin{document}
\maketitle

\section{Summary}\label{summary}

The Python Battery Optimisation and Parameterisation (\texttt{PyBOP})
package provides methods for estimating and optimising battery model
parameters, offering both deterministic and stochastic approaches with
example workflows to assist users. \texttt{PyBOP} enables parameter
identification from data for various battery models, including the
electrochemical and equivalent circuit models provided by the popular
open-source \texttt{PyBaMM} package (\citeproc{ref-Sulzer:2021}{Sulzer
et al., 2021}). Using the same approaches, \texttt{PyBOP} can also be
used for design optimisation under user-defined operating conditions
across a variety of model structures and design goals. \texttt{PyBOP}
facilitates optimisation with a range of methods, with diagnostics for
examining optimiser performance and convergence of the cost and
corresponding parameters. Identified parameters can be used for
prediction, on-line estimation and control, and design optimisation,
accelerating battery research and development.

\section{Statement of need}\label{statement-of-need}

\texttt{PyBOP} is a Python package providing a user-friendly,
object-oriented interface for optimising battery model parameters.
\texttt{PyBOP} leverages the open-source \texttt{PyBaMM} package
(\citeproc{ref-Sulzer:2021}{Sulzer et al., 2021}) to formulate and solve
battery models. Together, these tools serve a broad audience including
students, engineers, and researchers in academia and industry, enabling
the use of advanced models where previously this was not possible
without specialised knowledge of battery modelling, parameter inference,
and software development. \texttt{PyBOP} emphasises clear and
informative diagnostics and workflows to support users with varying
levels of domain expertise, and provides access to a wide range of
optimisation and sampling algorithms. These are enabled through
interfaces to \texttt{PINTS} (\citeproc{ref-Clerx:2019}{Clerx et al.,
2019}), \texttt{SciPy} (\citeproc{ref-SciPy:2020}{Virtanen et al.,
2020}), and \texttt{PyBOP}'s own implementations of algorithms such as
adaptive moment estimation with weight decay (AdamW)
(\citeproc{ref-Loshchilov:2017}{Loshchilov \& Hutter, 2017}), gradient
descent (\citeproc{ref-Cauchy:1847}{Cauchy \& others, 1847}), and cuckoo
search (\citeproc{ref-Yang:2009}{Yang \& Suash Deb, 2009}).

\texttt{PyBOP} supports the battery parameter exchange (BPX) standard
(\citeproc{ref-BPX:2023}{Korotkin et al., 2023}) for sharing parameter
sets. These are typically costly to obtain due to the specialised
equipment and time required for characterisation experiments, the need
for domain knowledge, and the computational cost of estimation.
\texttt{PyBOP} reduces the requirements for the latter two by providing
fast parameter estimation methods, standardised workflows, and parameter
set interoperability (via BPX).

\texttt{PyBOP} complements other lithium-ion battery modelling packages
built around \texttt{PyBaMM}, such as \texttt{liionpack} for battery
pack simulation (\citeproc{ref-Tranter2022}{Tranter et al., 2022}) and
\texttt{pybamm-eis} for fast numerical computation of the
electrochemical impedance of any battery model. Identified
\texttt{PyBOP} parameters are easily exportable to other packages.

\section{Architecture}\label{architecture}

\texttt{PyBOP} has a layered structure enabling the necessary
functionality to compute forward predictions, process results, and run
optimisation and sampling algorithms. The forward model is solved using
the battery modelling software \texttt{PyBaMM}, with construction,
parameterisation, and discretisation managed by \texttt{PyBOP}'s model
interface to \texttt{PyBaMM}. This provides a robust object construction
process with a consistent interface between forward models and
optimisers. Furthermore, identifiability metrics are provided along with
the estimated parameters (through Hessian approximation of the cost
functions around the optimum point in frequentist workflows, and
posterior distributions in Bayesian workflows).\\

\begin{figure}
\centering
\includegraphics[width=0.85\textwidth,height=\textheight]{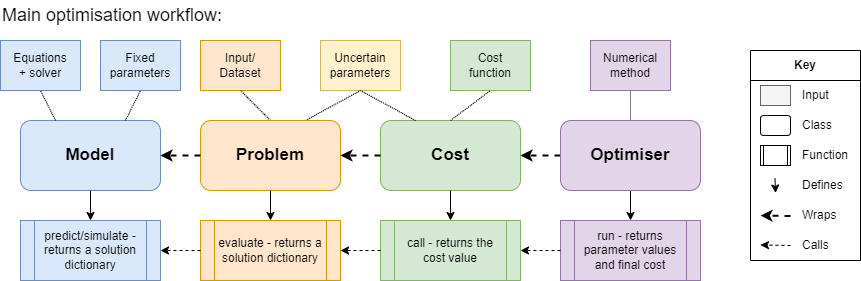}
\caption{The core \texttt{PyBOP} architecture with base class
interfaces. Each class provides a direct mapping to a step in the
optimisation workflow. \label{fig:classes}}
\end{figure}
\vspace{0.5em}

\texttt{PyBOP} formulates the inference process into four key classes:
model, problem, cost (or likelihood), and optimiser (or sampler), as
shown in \autoref{fig:classes}. Each of these objects represents a base
class with child classes constructing specialised functionality for
different workflows. The model class constructs a \texttt{PyBaMM}
forward model with a specified set of equations, initial conditions,
spatial discretisation, and numerical solver. By composing
\texttt{PyBaMM} directly into \texttt{PyBOP}, specialised models can be
constructed alongside the standard models that can also be modified for
different inference tasks. One such example is spatial
re-discretisation, which is required when one or more geometric
parameters are being optimised. In this situation, \texttt{PyBOP}
rebuilds the \texttt{PyBaMM} model only when necessary, reducing the
total number of rebuilds, providing improved performance. Alongside
construction of the forward model, \texttt{PyBOP}'s model class provides
methods for obtaining sensitivities from the prediction, enabling
gradient-based optimisation. A forward prediction, along with its
corresponding sensitivities, is provided to the problem class for
processing and exception control. A standardised data structure is then
provided to the cost classes, which computes a distance, design, or
likelihood-based metric for optimisation. For point-based optimisation,
the optimisers minimise the cost function or the negative log-likelihood
if a likelihood class is provided. Bayesian inference is provided by
sampler classes, which accept the \texttt{LogPosterior} class and sample
from it using \texttt{PINTS}-based Monte Carlo algorithms at the time of
submission. In the typical workflow, the classes in
\autoref{fig:classes} are constructed in sequence, from left to right in
the figure.

In addition to the core architecture, \texttt{PyBOP} provides several
specialised inference and optimisation features. One example is
parameter inference from electrochemical impedance spectroscopy (EIS)
simulations, where \texttt{PyBOP} discretises and linearises the EIS forward
model into a sparse mass matrix form with accompanying
auto-differentiated Jacobian. This is then translated into the frequency
domain, giving a direct solution to compute the input-output impedance.
In this situation, the forward models are constructed within the spatial
re-discretisation workflow, allowing for geometric parameter inference
from EIS simulations and data.

A second specialised feature is that \texttt{PyBOP} builds on the
\texttt{JAX} (\citeproc{ref-jax:2018}{Bradbury et al., 2018}) numerical
solvers used by \texttt{PyBaMM} by providing \texttt{JAX}-based cost
functions for automatic forward model differentiation with respect to
the parameters. This functionality provides a performance improvement
and allows users to harness many other \texttt{JAX}-based inference packages to
optimise cost functions, such as \texttt{NumPyro}
(\citeproc{ref-numpyro:2019}{Phan et al., 2019}), \texttt{BlackJAX}
(\citeproc{ref-blackjax:2024}{Cabezas et al., 2024}), and \texttt{Optax}
(\citeproc{ref-optax:2020}{DeepMind et al., 2020}).

The currently implemented subclasses for the model, problem, and cost
classes are listed in \autoref{tab:subclasses}. The model and optimiser
classes can be selected in combination with any problem-cost pair.\\

\renewcommand*{\arraystretch}{1.4}
\begin{longtable}[]{@{}
  >{\raggedright\arraybackslash}p{(\columnwidth - 4\tabcolsep) * \real{0.4458}}
  >{\raggedright\arraybackslash}p{(\columnwidth - 4\tabcolsep) * \real{0.2048}}
  >{\raggedright\arraybackslash}p{(\columnwidth - 4\tabcolsep) * \real{0.3494}}@{}}
\caption{List of available model, problem and cost/likelihood
classes. \label{tab:subclasses}}\tabularnewline
\toprule\noalign{}
\begin{minipage}[b]{\linewidth}\raggedright
Battery Models
\end{minipage} & \begin{minipage}[b]{\linewidth}\raggedright
Problem Types
\end{minipage} & \begin{minipage}[b]{\linewidth}\raggedright
Cost / Likelihood 
\end{minipage} \\
\midrule\noalign{}
\endfirsthead
\toprule\noalign{}
\begin{minipage}[b]{\linewidth}\raggedright
Battery Models
\end{minipage} & \begin{minipage}[b]{\linewidth}\raggedright
Problem Types
\end{minipage} & \begin{minipage}[b]{\linewidth}\raggedright
Cost / Likelihood Functions
\end{minipage} \\
\midrule\noalign{}
\endhead
\bottomrule\noalign{}
\endlastfoot

Single-particle model (SPM) & Fitting problem & Sum-squared error \\
SPM with electrolyte (SPMe) & Design problem & Root-mean-squared
error \\
Doyle-Fuller-Newman (DFN) & Observer & Minkowski \\
Many-particle model (MPM) & & Sum-of-power \\
Multi-species multi-reaction (MSMR) & & Gaussian log likelihood \\
Weppner Huggins & & Maximum a posteriori \\
Equivalent circuit model (ECM) & & Volumetric energy density \\
& & Gravimetric energy density \\
\end{longtable}

Similarly, the current algorithms available for optimisation are
presented in \autoref{tab:subclasses}. It should be noted that \texttt{SciPy}
minimize includes several gradient and non-gradient methods. From here
on, the point-based parameterisation and design-optimisation tasks will
simply be referred to as optimisation tasks. This simplification can be
justified by comparing \autoref{eqn:parameterisation} and
\autoref{eqn:design}; deterministic parameterisation is just an
optimisation task to minimise a distance-based cost between model output
and measured values.\\

\begin{longtable}[]{@{}
  >{\raggedright\arraybackslash}p{4.5cm}
  >{\raggedright\arraybackslash}p{4cm}
  >{\raggedright\arraybackslash}p{3.5cm}@{}}
\caption{Currently supported optimisers classified by candidate solution
type, including gradient information.
\label{tab:optimisers}}\tabularnewline
\toprule\noalign{}
\begin{minipage}[b]{\linewidth}\raggedright
Gradient-based
\end{minipage} & \begin{minipage}[b]{\linewidth}\raggedright
Evolutionary
\end{minipage} & \begin{minipage}[b]{\linewidth}\raggedright
(Meta)heuristic
\end{minipage} \\
\midrule\noalign{}
\endfirsthead
\toprule\noalign{}
\begin{minipage}[b]{\linewidth}\raggedright
Gradient-based
\end{minipage} & \begin{minipage}[b]{\linewidth}\raggedright
Evolutionary
\end{minipage} & \begin{minipage}[b]{\linewidth}\raggedright
(Meta)heuristic
\end{minipage} \\
\midrule\noalign{}
\endhead
\bottomrule\noalign{}
\endlastfoot

Weight decayed adaptive moment estimation (AdamW) & Covariance matrix
adaptation (CMA-ES) & Particle swarm (PSO) \\
Improved resilient backpropagation (iRProp-) & Exponential natural
(xNES) & Nelder-Mead \\
Gradient descent & Separable natural (sNES) & Cuckoo search \\
SciPy minimize & Differential evolution & \\
\end{longtable}

In addition to deterministic optimisers \autoref{tab:subclasses},
\texttt{PyBOP} also provides Monte Carlo sampling routines to estimate
distributions of parameters within a Bayesian framework. These methods
construct a posterior parameter distribution that can be used to assess
uncertainty and practical identifiability. The individual sampler
classes are currently composed within \texttt{PyBOP} from the
\texttt{PINTS} library, with a base sampler class implemented for
interoperability and direct integration with \texttt{PyBOP}'s model,
problem, and likelihood classes. The currently supported samplers are
listed in \autoref{tab:samplers}.\\

\begin{longtable}[]{@{}
  >{\raggedright\arraybackslash}p{2.5cm}
  >{\raggedright\arraybackslash}p{3cm}
  >{\raggedright\arraybackslash}p{2.3cm}
  >{\raggedright\arraybackslash}p{2.3cm}
  >{\raggedright\arraybackslash}p{3cm}@{}}
\caption{Sampling methods supported by \texttt{PyBOP}, classified
according to the candidate proposal method. \label{tab:samplers}}\tabularnewline
\toprule\noalign{}
\begin{minipage}[b]{\linewidth}\raggedright
Gradient-based
\end{minipage} & \begin{minipage}[b]{\linewidth}\raggedright
Adaptive
\end{minipage} & \begin{minipage}[b]{\linewidth}\raggedright
Slicing
\end{minipage} & \begin{minipage}[b]{\linewidth}\raggedright
Evolutionary
\end{minipage} & \begin{minipage}[b]{\linewidth}\raggedright
Other
\end{minipage} \\
\midrule\noalign{}
\endfirsthead
\toprule\noalign{}
\begin{minipage}[b]{\linewidth}\raggedright
Gradient-based
\end{minipage} & \begin{minipage}[b]{\linewidth}\raggedright
Adaptive
\end{minipage} & \begin{minipage}[b]{\linewidth}\raggedright
Slicing
\end{minipage} & \begin{minipage}[b]{\linewidth}\raggedright
Evolutionary
\end{minipage} & \begin{minipage}[b]{\linewidth}\raggedright
Other
\end{minipage} \\
\midrule\noalign{}
\endhead
\bottomrule\noalign{}
\endlastfoot
Monomial gamma & Delayed rejection adaptive & Rank shrinking &
Differential evolution & Metropolis random walk \\
No-U-turn & Haario Bardenet & Doubling & & Emcee hammer \\
Hamiltonian & Haario & Stepout & & Metropolis adjusted Langevin \\
Relativistic & Rao Blackwell & & & \\
\end{longtable}

\section{Background}\label{background}

\subsection{Battery models}\label{battery-models}

In general, battery models, after spatial discretisation, can be written
in the form of a differential-algebraic system of equations,
\begin{equation}
\frac{\mathrm{d} \mathbf{x}}{\mathrm{d} t} = f(t,\mathbf{x},\boldsymbol{\theta}),
\label{dynamics}
\end{equation} \begin{equation}
0 = g(t, \mathbf{x}, \boldsymbol{\theta}),
\label{algebraic}
\end{equation} \begin{equation}
\mathbf{y}(t) = h(t, \mathbf{x}, \boldsymbol{\theta}),
\label{output}
\end{equation} with initial conditions \begin{equation}
\mathbf{x}(0) = \mathbf{x}_0(\boldsymbol{\theta}).
\label{initial_conditions}
\end{equation}

Here, \(t\) is time, \(\mathbf{x}(t)\) are the spatially discretised
states, \(\mathbf{y}(t)\) are the outputs (e.g.~the terminal voltage)
and \(\boldsymbol{\theta}\) are the unknown parameters.

Common battery models include various types of equivalent circuit models
(e.g.~the Thévenin model), the Doyle--Fuller--Newman (DFN) model
(\citeproc{ref-Doyle:1993}{Doyle et al., 1993};
\citeproc{ref-Fuller:1994}{Fuller et al., 1994}) based on porous
electrode theory, and its reduced-order variants including the single
particle model (SPM) (\citeproc{ref-Planella:2022}{Brosa Planella et
al., 2022}) and the multi-species multi-reaction (MSMR) model
(\citeproc{ref-Verbrugge:2017}{Verbrugge et al., 2017}). Simplified
models that retain acceptable predictive accuracy at lower computational
cost are widely used, for example in battery management systems, while
physics-based models are required to understand the impact of physical
parameters on performance. This separation of complexity traditionally
results in multiple parameterisations for a single battery type,
depending on the model structure.

\section{Examples}\label{examples}

\subsection{Parameterisation}\label{parameterisation}

The parameterisation of battery models is challenging due to the large
number of parameters that need to be identified compared to the number
of measurable outputs (\citeproc{ref-Andersson:2022}{Andersson et al.,
2022}; \citeproc{ref-Miguel:2021}{Miguel et al., 2021};
\citeproc{ref-Wang:2022}{Wang et al., 2022}). A complete
parameterisation often requires stepwise identification of smaller sets
of parameters from a variety of excitations and different data sets
(\citeproc{ref-Chen:2020}{Chen et al., 2020};
\citeproc{ref-Chu:2019}{Chu et al., 2019}; \citeproc{ref-Kirk:2022}{Kirk
et al., 2023}; \citeproc{ref-Lu:2021}{Lu et al., 2021}). Furthermore,
parameter identifiability can be poor for a given set of excitations and
data sets, requiring improved experimental design in addition to
uncertainty capable identification methods
(\citeproc{ref-Aitio:2020}{Aitio et al., 2020}).

A generic data-fitting optimisation problem may be formulated as:
\begin{equation}
\min_{\boldsymbol{\theta}} ~ \mathcal{L}_{(\hat{\mathbf{y}_i})}(\boldsymbol{\theta}) ~~~
\textrm{subject to equations (\ref{dynamics})\textrm{-}(\ref{initial_conditions})}
\label{eqn:parameterisation}
\end{equation}

where \(\mathcal{L} : \boldsymbol{\theta} \mapsto [0,\infty)\) is a cost
function that quantifies the agreement between the model output
\(\mathbf{y}(t)\) and a sequence of observations
\((\hat{\mathbf{y}_i})\) measured at times \(t_i\). Within the
\texttt{PyBOP} framework, the \texttt{FittingProblem} class packages the
model output along with the measured observations, both of which are
then passed to the cost classes for the computation of the specific cost
function. For gradient-based optimisers, the Jacobian of the cost
function with respect to unknown parameters,
\(\partial \mathcal{L} / \partial \theta\), is computed for step-size
and directional information.

\begin{figure}[H]%[htb]
\centering
\includegraphics[width=1\textwidth,height=\textheight]{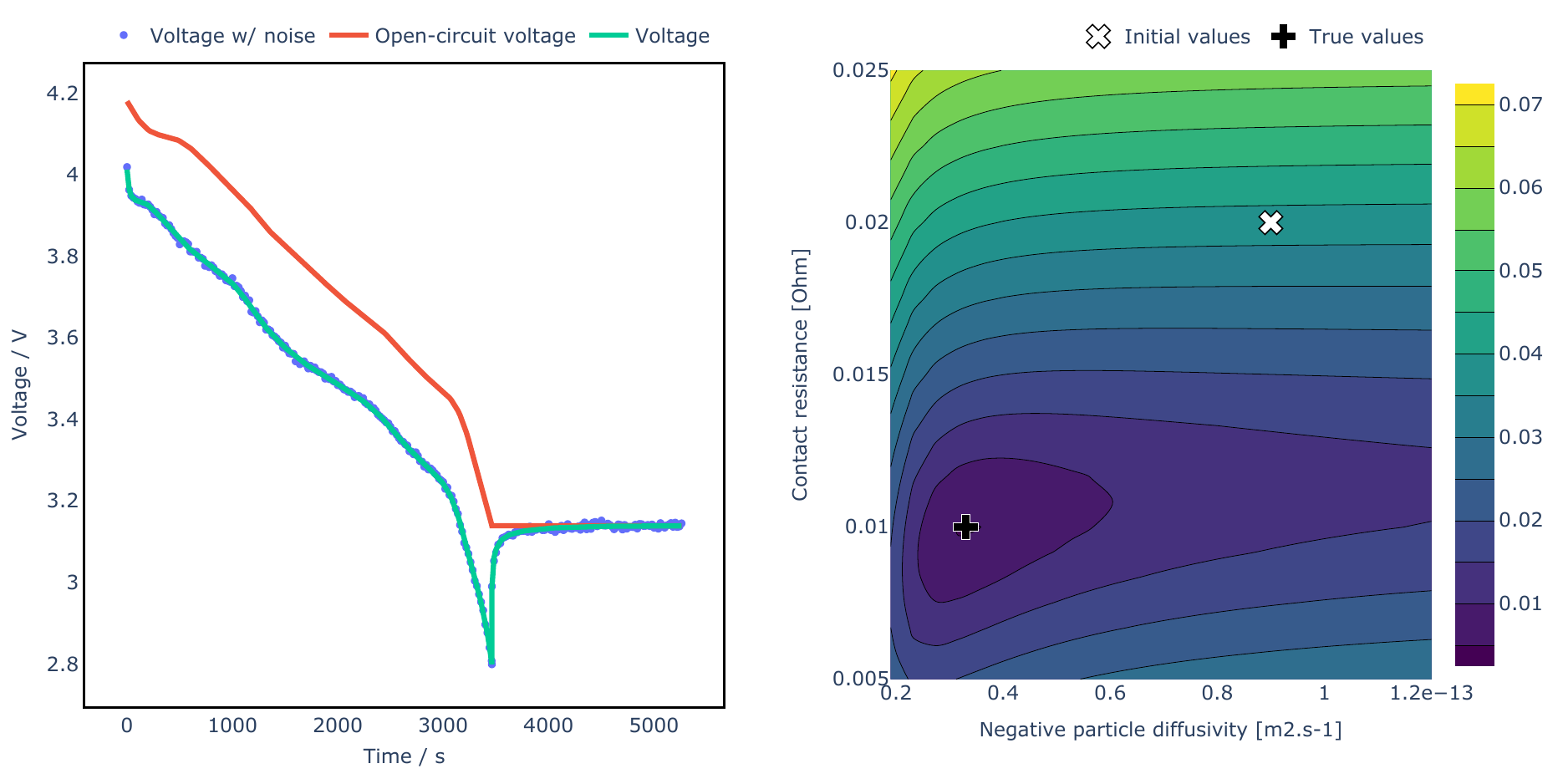}
\caption{The synthetic fitting dataset (left) and cost landscape (right)
for an example time-series battery model parameterisation using a
root-mean-squared error cost function.
\label{fig:inference-time-landscape}}
\end{figure}

Next, we demonstrate the fitting of synthetic data where the model
parameters are known. Throughout this section, as an example, we use
\texttt{PyBaMM}'s implementation of the single particle model with an
added contact resistance submodel. We assume that the model is already
fully parameterised apart from two dynamic parameters, namely, the
lithium diffusivity of the negative electrode active material particles
(denoted ``negative particle diffusivity'') and the contact resistance
with corresponding true values of {[}3.3e-14 \(\text{m}^2/\text{s}\), 10
mOhm{]}. To start, we generate synthetic time-domain data correspondinog to a one-hour discharge from 100\%
to 0\% state of charge, denoted as 1C rate, followed by 30 minutes of
relaxation. This dataset is then corrupted with zero-mean Gaussian noise
of amplitude 2 mV, with the resulting signal shown by the blue dots in
\autoref{fig:inference-time-landscape} (left). The initial states are
assumed known, although this assumption is not generally necessary. The
\texttt{PyBOP} repository contains several other
\href{https://github.com/pybop-team/PyBOP/tree/develop/examples/notebooks}{example
notebooks} that follow a similar inference process. The underlying cost
landscape to be explored by the optimiser is shown in
\autoref{fig:inference-time-landscape} (right), with the initial
position denoted alongside the known true system parameters for this
synthetic inference task. In general, the true parameters are not known.

\begin{figure}[H]%htb]
\centering
\includegraphics[width=1\textwidth,height=\textheight]{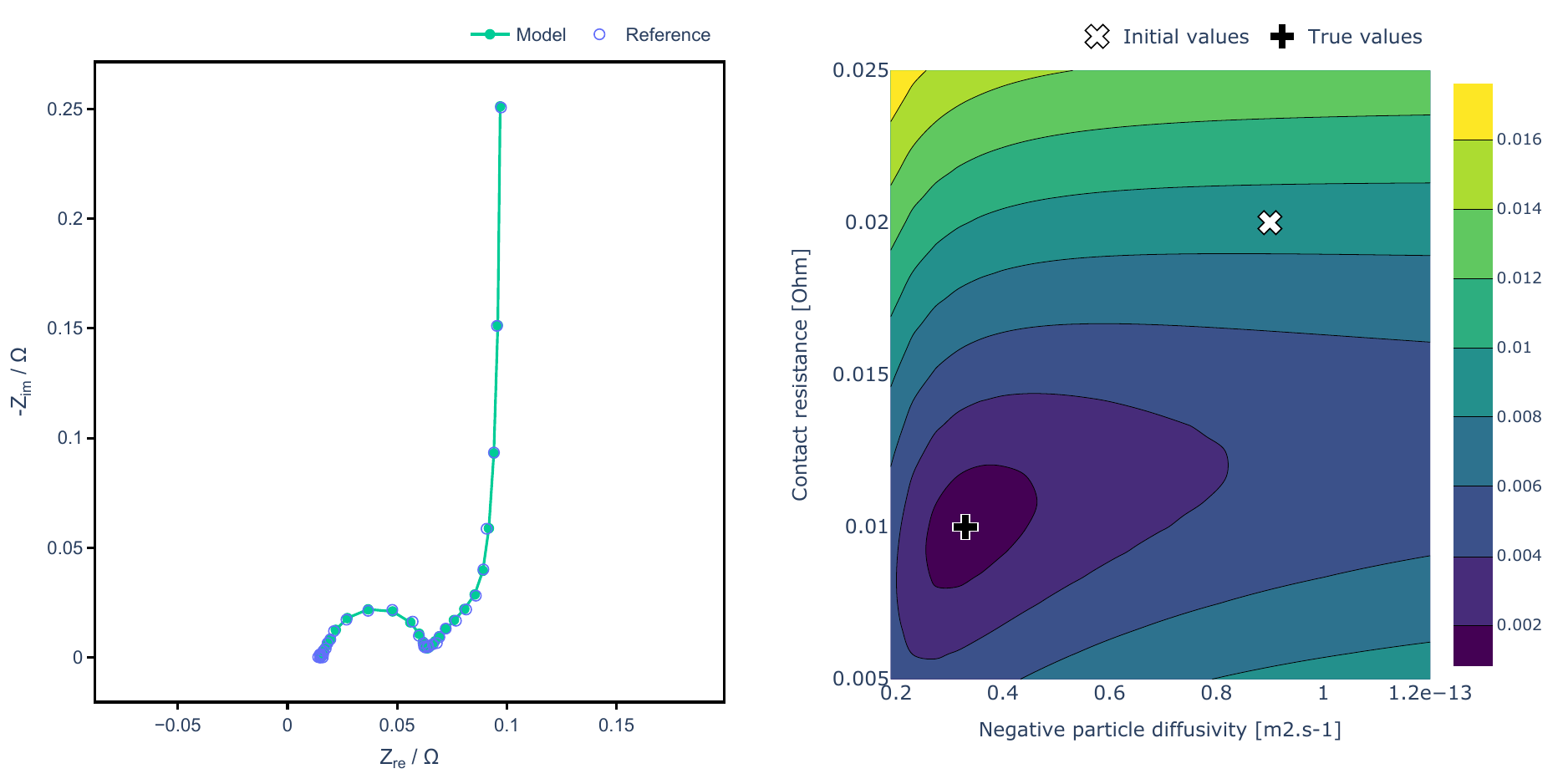}
\caption{The data and model fit (left) and cost landscape (right) for a
frequency-domain impedance parameterisation with a root-mean-squared
error cost function, at 5\% SOC. \label{fig:impedance-landscape}}
\end{figure}

We can also use \texttt{PyBOP}'sto generate and fit electrochemical 
impedance data using methods within 
\texttt{pybamm-eis} that enable fast impedance computation of battery
models (\citeproc{ref-pybamm-eis}{Dhoot et al., 2024}).Using the same model and parameters as in the time-domain case, \autoref{fig:impedance-landscape} 
shows the numerical impedance prediction available in \texttt{PyBOP} 
alongside the cost landscape for the corresponding inference task. At the 
time of publication, gradient-based optimisation and sampling methods are 
not available when using an impedance workflow.

\begin{figure}[H]%[htb]
\centering
\includegraphics[width=1\textwidth,height=\textheight]{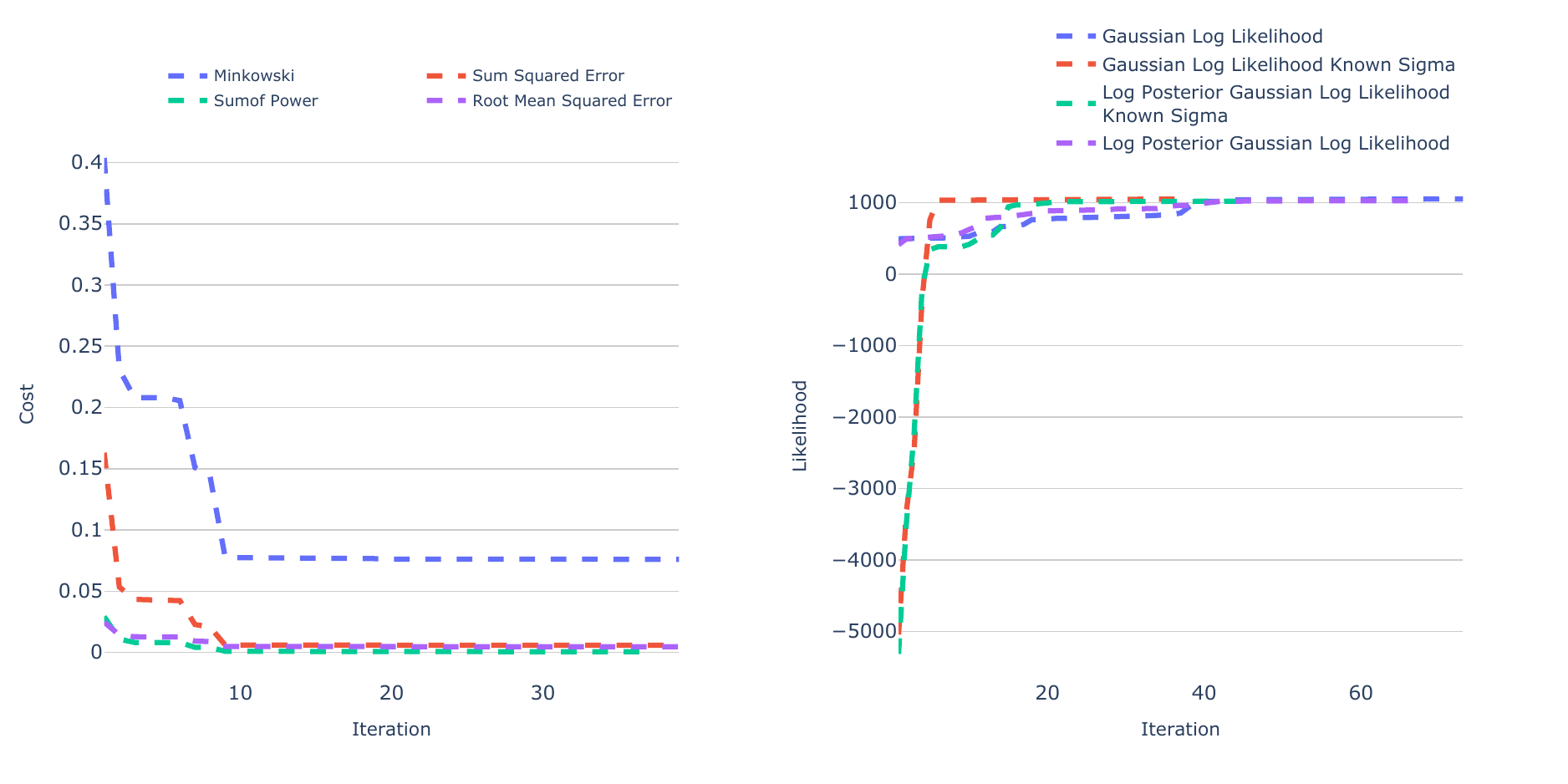}
\caption{Optimiser convergence using various cost (left) and likelihood
(right) functions and the L-BFGS-B algorithm.
\label{fig:convergence-min-max}}
\end{figure}

To avoid confusion, we continue
with identification in the time-domain (\autoref{fig:inference-time-landscape}). In general, however, time- and frequency-domain models and data may be combined for
improved parameterisation. As gradient information is available for our
time-domain example, the choice of distance-based cost function and
optimiser is not constrained. Due to the difference in magnitude between
the two parameters, we apply the logarithmic parameter transformation
offered by \texttt{PyBOP}. This transforms the search space of the
optimiser to allow for a common step size between the parameters,
improving convergence in this particular case. As a demonstration of the
parameterisation capabilities of \texttt{PyBOP},
\autoref{fig:convergence-min-max} (left) shows the rate of convergence
for each of the distance-minimising cost functions, while
\autoref{fig:convergence-min-max} (right) shows analogous results for
maximising a likelihood. The optimisation is performed with \texttt{SciPy}
Minimize using the gradient-based L-BFGS-B method.

\begin{figure}[H]
\centering
\includegraphics[width=1\textwidth,height=\textheight]{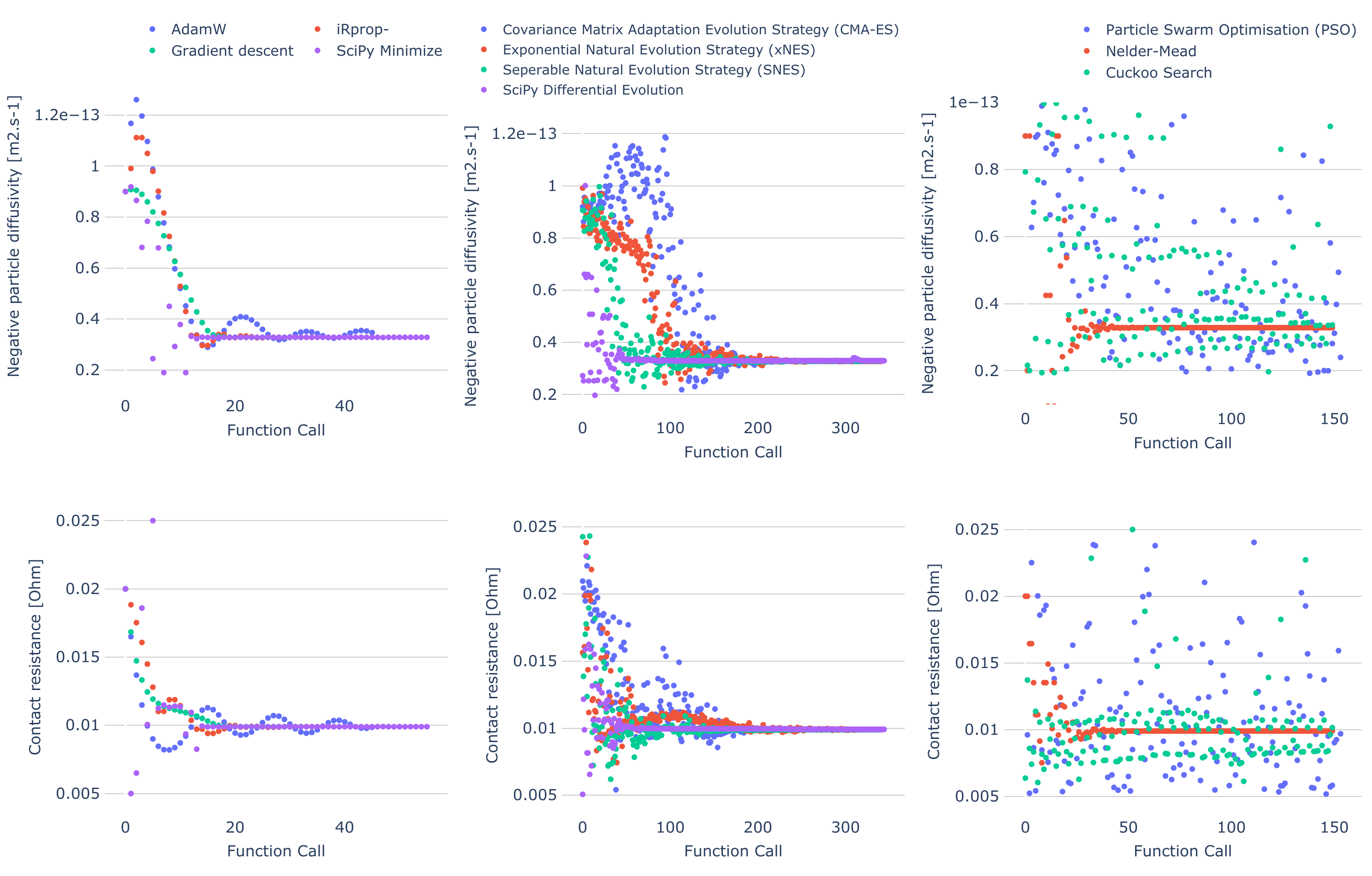}
\caption{Convergence in parameter values for several optimisation
algorithms provided by \texttt{PyBOP}. \label{fig:optimiser-inference1}}
\end{figure}

Using the same model and parameters, we compare example convergence 
rates of various algorithms across several categories:
gradient-based methods in \autoref{fig:optimiser-inference1} (left),
evolutionary strategies in \autoref{fig:optimiser-inference1} (middle)
and (meta)heuristics in \autoref{fig:optimiser-inference1} (right) using
a mean-squared-error cost function. \\

\begin{figure}[H]
\centering
\includegraphics[width=1\textwidth,height=\textheight]{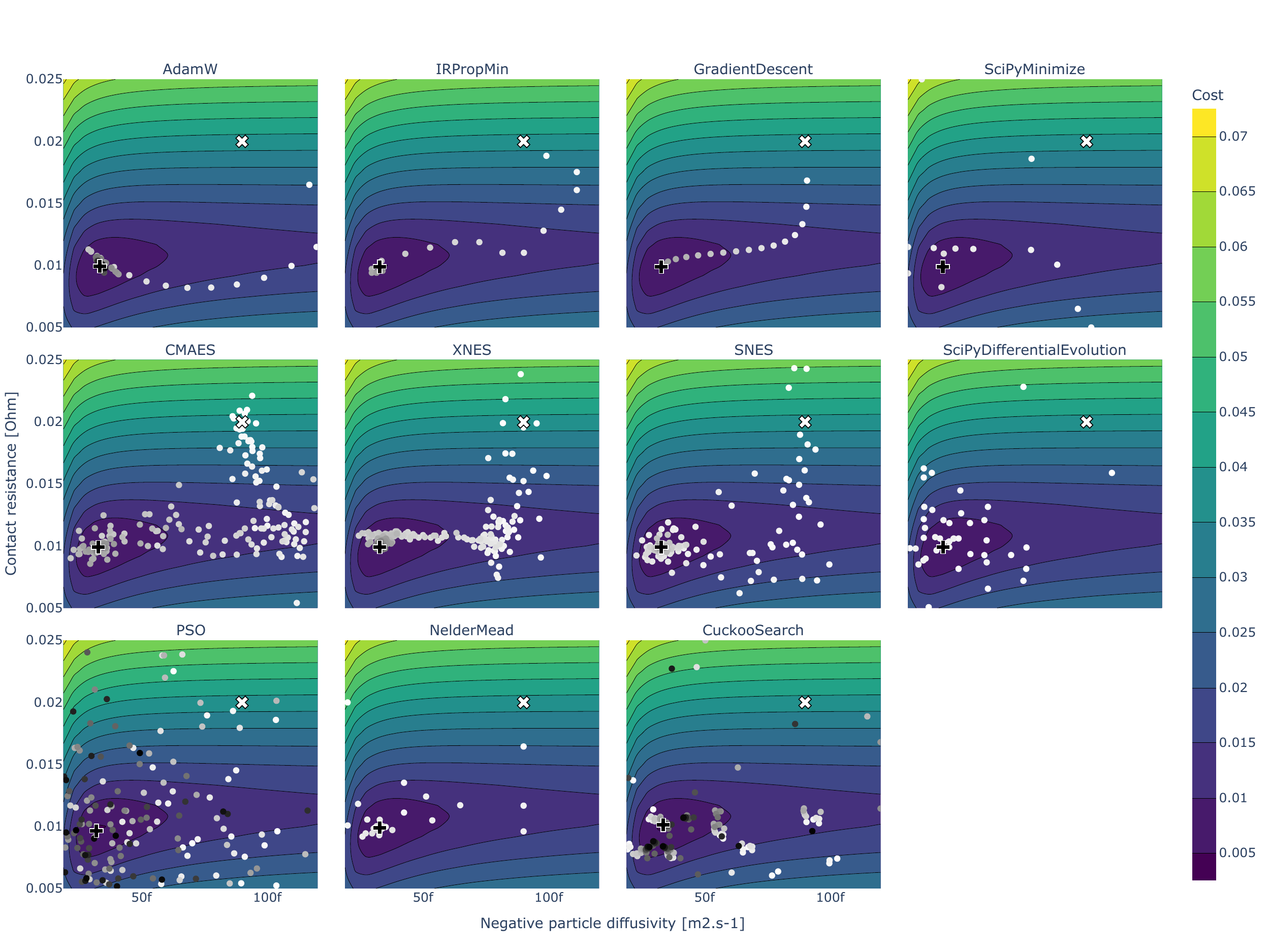}
\caption{Cost landscape contour plot with corresponding optimisation
traces, for several optimisers. \label{fig:optimiser-inference2}}
\end{figure}

We also show the cost function landscapes alongside optimiser iterations in \autoref{fig:optimiser-inference2}, 
with the three rows showing the gradient-based optimisers (top), evolution
strategies (middle), and (meta)heuristics (bottom). Note that the
performance of the optimiser depends on the cost landscape, the initial
guess or prior, and the hyperparameters for each specific
problem.

This example parameterisation task can also be approached from a
Bayesian perspective, using \texttt{PyBOP}`s sampler methods.
First, we introduce Bayes' rule,

\begin{equation}
P(\theta|D) = \frac{P(D|\theta)P(\theta)}{P(D)},
\label{eqn:bayes_theorem}
\end{equation}
\vspace{0.5em}

where \(P(\theta|D)\) is the posterior parameter distribution,
\(P(D|\theta)\) is the likelihood function, \(P(\theta)\) is the prior
parameter distribution, and \(P(D)\) is the model evidence, or marginal
likelihood, which acts as a normalising constant. In the case of maximum
likelihood estimation or maximum a posteriori estimation, one wishes to
maximise \(P(D|\theta)\) or \(P(\theta|D)\), respectively, and this may
be formulated as an optimisation problem as per
\autoref{eqn:parameterisation}.

\begin{figure}[htb]
\centering
\includegraphics[width=1\textwidth,height=\textheight]{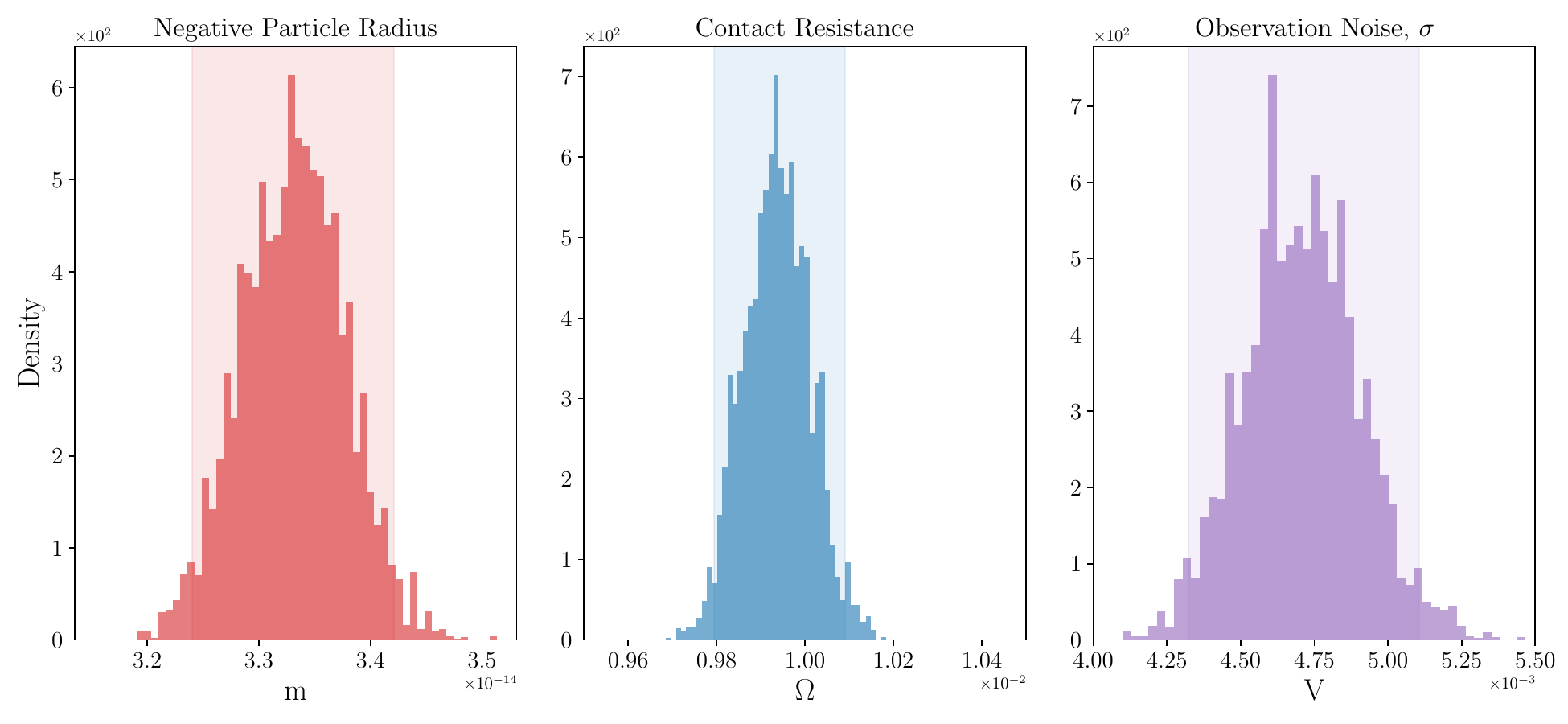}
\caption{Posterior distributions of model parameters alongside
identified noise on the observations. Shaded areas denote the 95th
percentile credible interval for each parameter. \label{fig:posteriors}}
\end{figure}

One must use sampling or other inference methods to reconstruct the full posterior
parameter distribution, \(P(\theta|D)\). The posterior distribution provides information about
the uncertainty of the identified parameters, e.g., by calculating the
variance or other moments. Monte Carlo methods are used here to sample
from the posterior. The selection of Monte Carlo methods available in
\texttt{PyBOP} includes gradient-based methods such as no-u-turn
(\citeproc{ref-NUTS:2011}{Hoffman \& Gelman, 2011}) and Hamiltonian
(\citeproc{ref-Hamiltonian:2011}{Brooks et al., 2011}), as well as
heuristic methods such as differential evolution
(\citeproc{ref-DiffEvolution:2006}{Braak, 2006}), and also conventional
methods based on random sampling with rejection criteria
(\citeproc{ref-metropolis:1953}{Metropolis et al., 1953}).
\texttt{PyBOP} offers a sampler class that provides the interface to
samplers, the latter being provided by the probabilistic inference on
noisy time-series (\texttt{PINTS}) package. \autoref{fig:posteriors}
shows the sampled posteriors for the synthetic model described
previously, using an adaptive covariance-based sampler called Haario
Bardenet (\citeproc{ref-Haario:2001}{Haario et al., 2001}).

\subsection{Design optimisation}\label{design-optimisation}

Design optimisation is supported in \texttt{PyBOP} to guide device
design development by identifying parameter sensitivities that can
unlock improvements in performance. This problem can be viewed in a
similar way to the parameterisation workflows described previously, but
with the aim of maximising a design-objective cost function rather than
minimising a distance-based cost function. \texttt{PyBOP} performs
maximisation by minimising the negative of the cost function. In design
problems, the cost metric is no longer a distance between two time
series, but a metric evaluated on a model prediction. For example, to
maximise the gravimetric energy (or power) density, the cost is the
integral of the discharge energy (or power) normalised by the cell mass.
Such metrics are typically quantified for operating conditions such as a
1C discharge, at a given temperature. In general, design optimisation can be written as a constrained
optimisation problem,

 \begin{equation}
\min_{\boldsymbol{\theta} \in \Omega} ~ \mathcal{L}(\boldsymbol{\theta}) ~~~
\textrm{subject to equations (\ref{dynamics})\textrm{-}(\ref{initial_conditions}),}
\label{eqn:design}
\end{equation}
\vspace{0.5em}

where \(\mathcal{L} : \boldsymbol{\theta} \mapsto [0,\infty)\) is a cost
function that quantifies the desirability of the design and \(\Omega\)
is the set of allowable parameter values.

As an example, we consider the challenge of maximising the gravimetric
energy density, subject to constraints on two of the geometric electrode
parameters (\citeproc{ref-Couto:2023}{Couto et al., 2023}). In this case
we use the \texttt{PyBaMM} implementation of the single particle model with electrolyte (SPMe) to investigate the impact of the positive electrode thickness and the active material volume fraction on the energy density. Since the total volume fraction must sum to unity, the positive electrode porosity for each optimisation iteration is defined in relation to the active material volume fraction. It is also possible to update the 1C rate corresponding to the theoretical capacity for each iteration of the design.

\begin{figure}[htb]
\centering
\includegraphics[width=1\textwidth,height=\textheight]{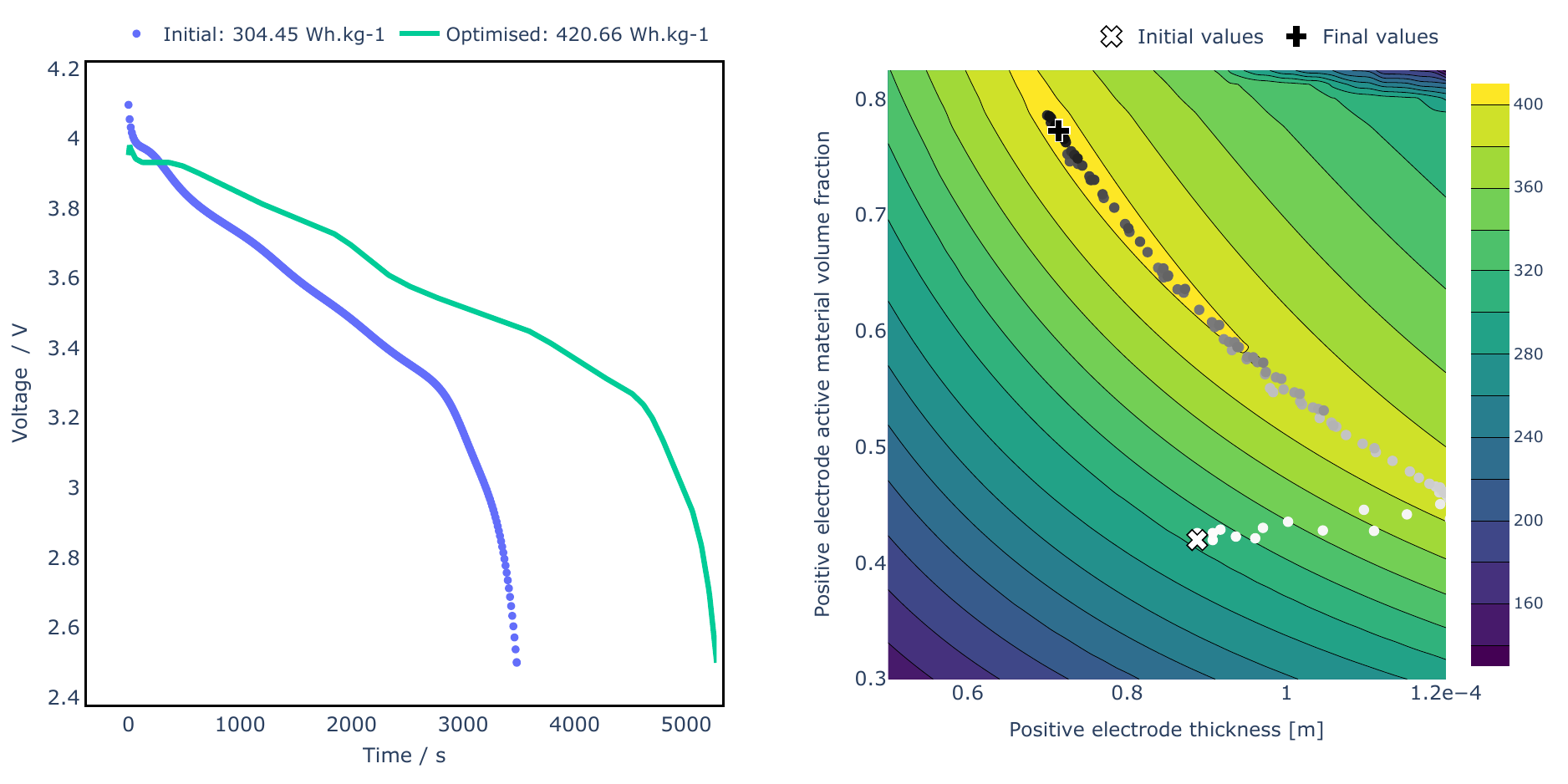}
\caption{Initial and optimised voltage profiles alongside the gravimetric energy density cost landscape. \label{fig:design_gravimetric}}
\end{figure}

\autoref{fig:design_gravimetric} (left) shows the predicted improvement in the discharge profile between the initial and optimised parameter values for a fixed-rate 1C discharge selected from the initial design and (right) the Nelder-Mead search over the parameter space.

\section{Conclusion}\label{conclusion}
In this article, the open-source parameter identification and optimisation package, \texttt{PyBOP}
has been introduced. This package provides specialised classes and methods to support both
physics-based and data-driven parameter identification and optimisation. An example identification workflow has been 
presented using both point-estimate and distribution-based inference methods for both frequency and time domain
forward model predictions. An additional workflow for design optimisation was also presented, showcasing \texttt{PyBOP} as a versatile package capable for researchers and engineers.

\section{Acknowledgements}\label{acknowledgements}

We gratefully acknowledge all
\href{https://github.com/pybop-team/PyBOP?tab=readme-ov-file\#contributors-}{contributors}
to \texttt{PyBOP}. This work was supported by the Faraday Institution
Multiscale Modelling project (ref. FIRG059), UKRI's Horizon Europe
Guarantee (ref. 10038031), and EU IntelLiGent project (ref. 101069765).

\section*{References}\label{references}
\addcontentsline{toc}{section}{References}

\phantomsection\label{refs}

\end{document}